# Renal function changes in chronic hepatitis B patients treated with pegylated interferon and oral antiviral drugs.

Jinhua Zhao[1]   Lili Wu[1,2]   Xiaoan Yang[1,2]   Zhiliang Gao[1,2]   Hong Deng[1]

[1]Department of Infectious Diseases, Third Affiliated Hospital of Sun Yat-sen University, Guangzhou, China

[2]Guangdong Provincial Key Laboratory of Liver Disease Research, Third Affiliated Hospital of Sun Yat-sen University, Guangzhou, China

**Correspondence**
Hong Deng, Department of Infectious Diseases, Third Affiliated Hospital of Sun Yat-sen University, 510630, Guangzhou, China.
Email: dhong@mail.sysu.edu.cn

Zhiliang Gao, Department of Infectious Diseases, The Third Affiliated Hospital of Sun Yat-sen University, No 600 Tianhe Road, Guangzhou, 510630, Guangdong Province, China;
Email: gaozhl@mail.sysu.edu.cn

Funding information
National Natural Science Foundation of China ,Grant Number: 81870597,82170612; Guangzhou Science and Technology Program Key Projects,Grant Number:2023B01J1007

**Abstract:**

Background and Aims:

The best way to treat chronic hepatitis B is with pegylated interferon alone or with oral antiviral drugs. There is limited research comparing the renal safety of entecavir and tenofovir when used with pegylated interferon. This study will compare changes in renal function in chronic hepatitis B patients treated with pegylated interferon and either entecavir or tenofovir.

Methods:

The study included a cohort of 836 patients with chronic hepatitis B (CHB) who received treatment with pegylated interferon (IFN) either alone or in combination with entecavir (ETV) and tenofovir (TDF) between the years 2018 and 2021. Of these patients, 713 were included in a matched analysis comparing outcomes between those who were cured and those who were uncured, while 123 patients received IFN alone as a control group for comparison with the ETV and TDF treatment groups. The primary outcome measured was the change in renal function, specifically estimated glomerular filtration rate (eGFR), cystatin C (CysC), and inorganic phosphorus (IPHOS). Patients were categorized into stage 1 or stage 2 based on a baseline eGFR of less than 90 ml/min/m².

Results:

125 CHB patients were matched 1:1 in both the combined treatment and cured groups. Baseline eGFR, CysC, and IPHOS levels were similar between the groups. Renal function in stage 1 and stage 2 groups showed a decreasing trend at 48 weeks after an initial increase. Correlation analysis showed significant relationships between changes in ALT and eGFR values at 12 weeks in both non-cured and cured groups. After matching, the study included 208 patients in the IFN plus ETV group, 208 patients in the IFN plus TDF group, and 104 patients in the IFN alone group with chronic hepatitis B. Baseline eGFR values were 103.44 ml/min/1.73m², 104.34 ml/min/1.73m², and 106.97 ml/min/1.73m², respectively. Mean baseline CysC levels were 0.87 mg/l, 0.88 mg/l, and 0.85 mg/l. Average baseline IPHOS levels were 1.04 mol/l, 1.05 mol/l, and 1.08 mol/l. No significant difference in the impact of three drug regimens on renal function at 48 weeks was found (P = 0.955). A logistic regression model using age and baseline eGFR predicted the AUC of eGFR status at 48 weeks to be 0.851 (95% CI, 0.807, 0.895).

Conclusions:

Over the 48-week duration of combined treatment in patients with chronic hepatitis B (CHB), it was found that both Tenofovir Disoproxil Fumarate (TDF) and Entecavir (ETV) did not lead to an increase in renal injury.

## Abbreviations:

PSM: propensity score matching
EMR:Electronic Medical Record
HBV:hepatitis B virus
K/DOQI:Kidney Disease Outcomes Quality Initiative
WBC:white blood cell counts
HGB: hemoglobin counts
PLT:platelets counts
LYM:lymphocyte count
NEUT:neutrophil neutrophil count
HBSAG:hepatitis B surface antigen
AST:aspartate aminotransferase
ALT:alanine aminotransferase

TBIL:total bilirubin serum total bilirubin
DBIL:direct bilirubin
ALB:albumin
eGFR: estimated glomerular filtration rate
IPHOS: inorganic phosphorus
CysC:serum cystatin C

**Key points:**

1、ETV and TDF did not impact renal function in CHB patients over 48 weeks when combined with pegylated interferon. TDF did not pose a higher risk of renal injury compared to ETV.
2、After 48 weeks of treatment, renal function improved in CHB patients who were cured compared to those who were uncured, possibly due to their response to pegylated interferon. However, the difference was not statistically significant, and both groups showed a similar trend of initially increasing and then decreasing renal function.
3、Age and baseline eGFR level predicted mild to moderate renal impairment in CHB patients after 48 weeks of treatment.Older patients or those with existing kidney issues should monitor their renal function closely when undergoing treatment with pegylated interferon or oral antiviral drugs.

**1 BACKGROUNND**

Chronic hepatitis B is a significant global health issue, with 3.8% of the population infected with HBsAg in 2019. This led to 1.5 million new infections, 296 million chronic cases, and 820,000 deaths from related liver diseases[1].Antiviral therapy is the main treatment for chronic hepatitis B, but it may not achieve HBsAg clearance, leading to lifelong treatment with drugs like ETV, TDF, TAF, and TMF[2,3,4].The latest guidelines recommend entecavir, tenofovir disoproxil fumarate, tenofovir alafenamide fumarate, and tenofovir amibufenamide as first-line drugs for CHB treatment. TDF may cause renal and bone safety issues, including hypophosphatemia, renal dysfunction, and osteomalacia[5].ETV has a good safety profile, with only 0.2% of patients experiencing serious adverse reactions in a 10-year global study[6].Multiple studies have indicated that patients with low HBsAg levels and negative HBeAg before interferon treatment are more likely to achieve clinical cure with sequential peg IFN- α treatment[7,8,9].Guidelines suggest using ETV or TAF as the preferred anti-HBV drugs for patients with kidney disease or receiving renal replacement therapy, and advise against using TDF[10,11].ETV is considered safer for the kidneys compared to TDF, making it the preferred choice for patients with reduced renal function[12].Research on renal function in CHB patients treated with pegylated interferon and ETV found that eGFR significantly improved within 48 weeks[13].No study has compared the effects on kidney function of pegylated interferon with ETV and TDF in CHB patients. This study compared renal function changes in patients treated with pegylated interferon or a combination of ETV and TDF.

## 2 METHODS

### 2.1 Study population

CHB patients treated with pegylated interferon or combined ETV and TDF between 2018 and 2020 were included in this retrospective study. All patients were part of a larger multicenter study in China[14]. The study followed clinical guidelines and ethical standards, approved by the medical ethics committee of the Third Affiliated Hospital of Sun Yat sen University, with written consent from all participants.

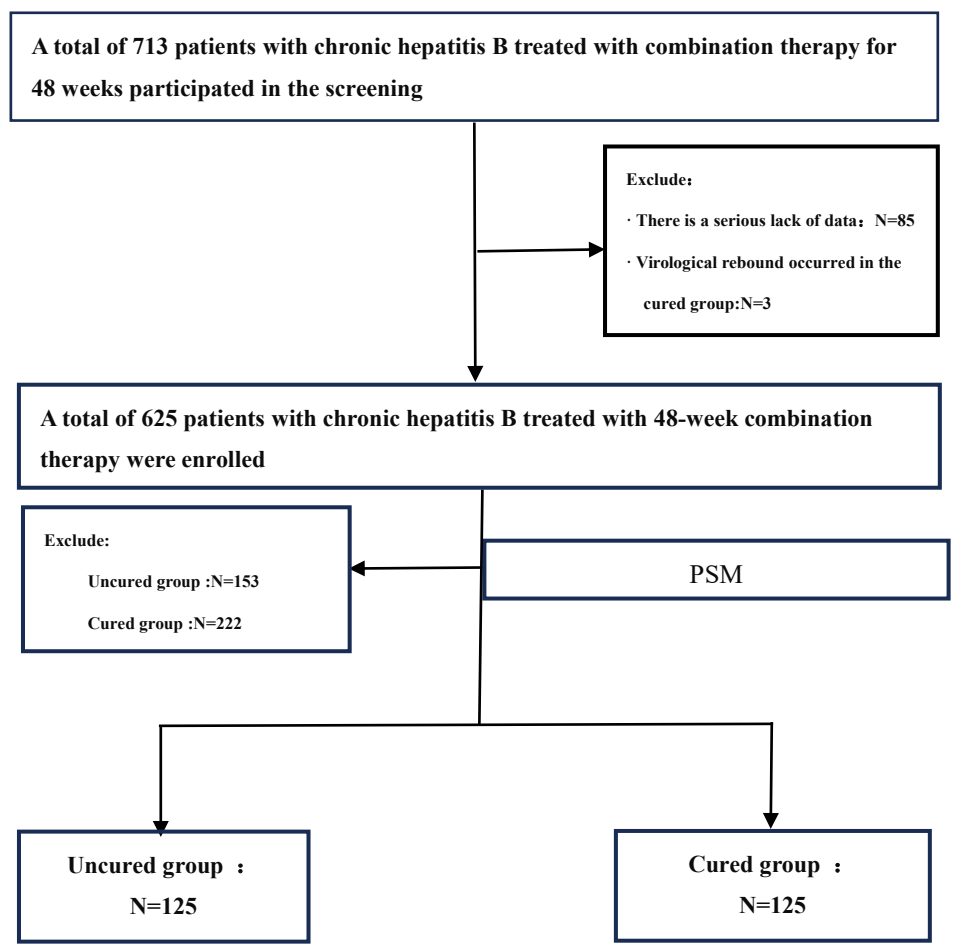

**FIGURE 1**. Flow chart of the study .PSM: propensity score matching.
**Note**:The included covariates were age, total number of pegylated interferon needles, HBsAg, eGFR0w

### 2.2 Data collection and follow-up

Since the start of AVT, demographic, treatment, and lab data of CHB patients were collected from the EMR system.Renal function indicators, such as eGFR, blood phosphorus, and blood cystatin C, were screened. eGFR was calculated using the CKD-EPI creatinine equation.CHB patients' detection index was measured at enrollment, with follow-up visits at weeks 0, 12, 24, 36, and 48 from 2018 to 2020.

2.3 Antiviral treatment

Throughout the study, patients were treated with either pegylated interferon alone or in combination with ETV and TDF from the start.Clinical cure was defined as sustained absence of HBsAg after treatment cessation, with or without anti-HBS, undetectable HBV DNA, normal liver enzymes, and possible presence of cccDNA in hepatocytes.

2.4 Outcome of renal function

The primary finding of this research was a slight decrease in estimated glomerular filtration rate (eGFR) after 48 weeks of observation. According to the guidelines set forth by the Kidney Disease Outcomes Quality Initiative (K/DOQI) established by the American Kidney Foundation, an eGFR of less than 90 ml/min/1.73m$^2$ is indicative of a mild reduction in kidney function (stage 1), while an eGFR of 90 ml/min/1.73m$^2$ or higher is considered normal (stage 2).

2.5 Statistical analysis

Missing values in SPSS were replaced with the sequence average, creating two complete datasets for maximum accuracy.To accurately elucidate the baseline characteristics of patients across various treatment status groups, we conducted propensity score matching at a ratio of 1:1 to equalize the baseline characteristics between patients in the cured and untreated groups. Additionally, a propensity score matching at a ratio of 2:2:1 was carried out to balance the baseline characteristics among patients receiving pegylated interferon alone, pegylated interferon combine with ETV, and pegylated interferon combine with TDF treatments.Matching method is nearest with a caliper value of 0.1, prioritizing complete matching. Continuous variables are shown as mean $\pm$ SD, and categorical variables are shown as number and percentage.Spearman correlation was used to analyze the relationship between ALT and eGFR changes. Multivariate analysis and ROC curve were used to predict 48-week renal function and assess the diagnostic value of predictors.Binary mixed logistic regression was utilized to predict 48-week renal function outcome, with statistical significance set at $P < 0.05$. IBM SPSS Statistics 27, GraphPad Prism 9, and web page analysis tools from https://www.xiantao.love were employed.

3 RESULTS

3.1 Baseline characteristics

Based on the inclusion criteria established for the enrolled patients, a total of 625 patients diagnosed with chronic hepatitis B (CHB) were selected for statistical analysis, comprising 278 individuals in the uncured group and 347 individuals in the cured group (Figure 1).The baseline characteristics of patients are presented in Table 1. Prior to matching, the average age of patients with chronic hepatitis B in the untreated group was notably higher (43.3 years vs. 39.1 years, P < 0.001).Following matching, there were no statistically significant differences observed in the average age and sex ratio between the chronic hepatitis B (CHB) patients in the uncured group and those in the cured group (P = 0.479, P = 0.054). Additionally, there were no statistically significant variances in the treatment regimen and baseline HBsAg level between the two groups (P = 0.513, P = 0.492).There was no statistically significant difference in baseline renal function between the two groups of chronic hepatitis B patients, as indicated by the levels of estimated glomerular filtration rate (eGFR), inorganic phosphorus (IPHOS), and cystatin C (CysC) (all P > 0.05).

TABLE1. Baseline characteristics of participant before and after PS matching.

| Baseline Characteristics | Before PSM (N=625) | | P Value | After PSM (N=250) | | P Value |
|---|---|---|---|---|---|---|
| | Uncured (N=278) | Cured (N=347) | | Uncured (N=125) | Cured (N=125) | |
| Age (year) | 43.35±8.66 | 39.15±8.52 | < 0.001 | 41.38±9.33 | 41.75±8.65 | 0.479 |
| Sex | | | | | | |
| male | 23 | 57 | < 0.001 | 113 | 110 | 0.054 |
| female | 255 | 290 | | 12 | 15 | |
| Treatment | | | | | | |
| The types of NAs in combination therapy | | | 0.341 | | | 0.513 |
| ETV | 135 (49%) | 162 (47%) | | 60 (48%) | 57 (46%) | |
| TDF | 110 (40%) | 149 (43%) | | 110 (40%) | 149 (43%) | |
| TAF | 20 (7%) | 15 (4%) | | 8 (6%) | 14 (10%) | |
| Other NAs | 13 (4%) | 21 (6%) | | 5 (4%) | 7 (6%) | |
| Number of IFN (n) | 53 | 45 | | 49 | 48 | |
| Blood routine | | | | | | |
| WBC (×$10^9$/L) | 6.13±1.48 | 5.88±1.56 | 0.058 | 6.07±1.33 | 5.87±1.56 | 0.291 |
| HGB (g/L) | 153.43±17.82 | 152.52±16.91 | 0.513 | 153.45±18.20 | 152.33±15.06 | 0.597 |
| PLT (×$10^9$/L) | 206.74±52.77 | 208.80±58.96 | 0.649 | 209.08±49.65 | 206.28±58.26 | 0.291 |
| LYM (×$10^9$/L) | 2.04±0.58 | 1.98±0.58 | 0.260 | 1.98±0.56 | 2.00±0.57 | 0.807 |
| NEUT (×$10^9$/L) | 3.48±1.19 | 3.27±1.23 | 0.026 | 3.47±1.07 | 3.25±1.21 | 0.122 |
| HBsAg (iu/ml) | 751.01±643.79 | 243.58±283.47 | < 0.001 | 454.42±288.32 | 426.81±343.72 | 0.492 |
| Blood biochemistry data | | | | | | |
| AST (U/L) | 31.52±18.67 | 31.07±25.92 | 0.811 | 29.82±16.17 | 32.54±31.26 | 0.388 |
| ALT (U/L) | 36.15±25.84 | 36.80±37.17 | 0.805 | 34.65±24.62 | 39.49±48.62 | 0.322 |
| TBIL (μmol/L) | 12.21±6.73 | 11.65±5.04 | 0.235 | 12.96±7.58 | 11.96±5.73 | 0.243 |
| DBIL (μmol/L) | 3.65±3.76 | 3.24±1.66 | 0.072 | 4.41±5.29 | 3.33±1.79 | 0.032 |
| ALB (g/L) | 47.58±3.18 | 48.20±2.74 | 0.010 | 47.84±3.20 | 48.24±2.66 | 0.280 |
| eGFR (ml/min/1.73m²) | 101.64±14.70 | 104.64±16.11 | 0.016 | 103.07±14.18 | 102.70±16.10 | 0.850 |
| IPHOS (mmol/L) | 1.03±0.16 | 1.06±0.29 | 0.119 | 1.04±0.14 | 1.08±0.34 | 0.308 |
| Cysc (mg/L) | 0.90±0.14 | 0.87±0.14 | 0.005 | 0.89±0.12 | 0.88±0.13 | 0.564 |

Note: continuous variables are expressed as mean ± standard deviation (SD), and categorical variables are expressed as number and percentage (%).

Abbreviations: WBC, white blood cell counts; HGB, hemoglobin counts; PLT, platelets counts; Lym, lymphocyte count; Neut, neutrophil neutrophil count; HBsAg, hepatitis B surface antigen; Ast, aspartate aminotransferase; Alt, alanine aminotransferase; TBIL, total bilirubin serum total bilirubin; DBIL, direct bilirubin; ALB, albumin, albumin; eGFR, estimated glomerular filtration rate; IPHOS, iphos, serum inorganic phosphorus; CysC, Cystatin C, serum cystatin.

[a]1: 1 covariates in the matched model included age, total number of interferon needles, baseline HBsAg, baseline alt, and baseline eGFR.

[b]Data conforming to normal distribution were analyzed for differences (or Wilcoxon rank

sum test) and chi square test using two independent samples t test.

3.2 The renal function

The study population's eGFR levels, stratified by cure status, gender, and age, indicated a decrease at 12 weeks of treatment among women and patients under 40 years of age in the cure group.After 12 weeks of treatment, the eGFR in the uncured group exhibited a decline, whereas the eGFR in the cured group demonstrated a continuous increase until the 24-week mark. Subsequently, a statistically significant decrease was observed in both groups at the 48-week mark ($P < 0.05$, Figure 2A).During the 48-week treatment period, women exhibited a higher eGFR compared to men. Specifically, women experienced an increase in eGFR at 36 weeks, while men continued to experience a decline. The disparity between the two groups was found to be statistically significant at 48 weeks ($P < 0.001$, Figure 2B).IPHOS levels increased after initially decreasing from 0 to 12 weeks in the cured state, gender, and age groups.CysC levels increased up to 36 weeks and then decreased in both groups, with no significant difference between them.CysC levels were higher in males over 40 years old compared to other groups at 48 weeks ($P < 0.05$).

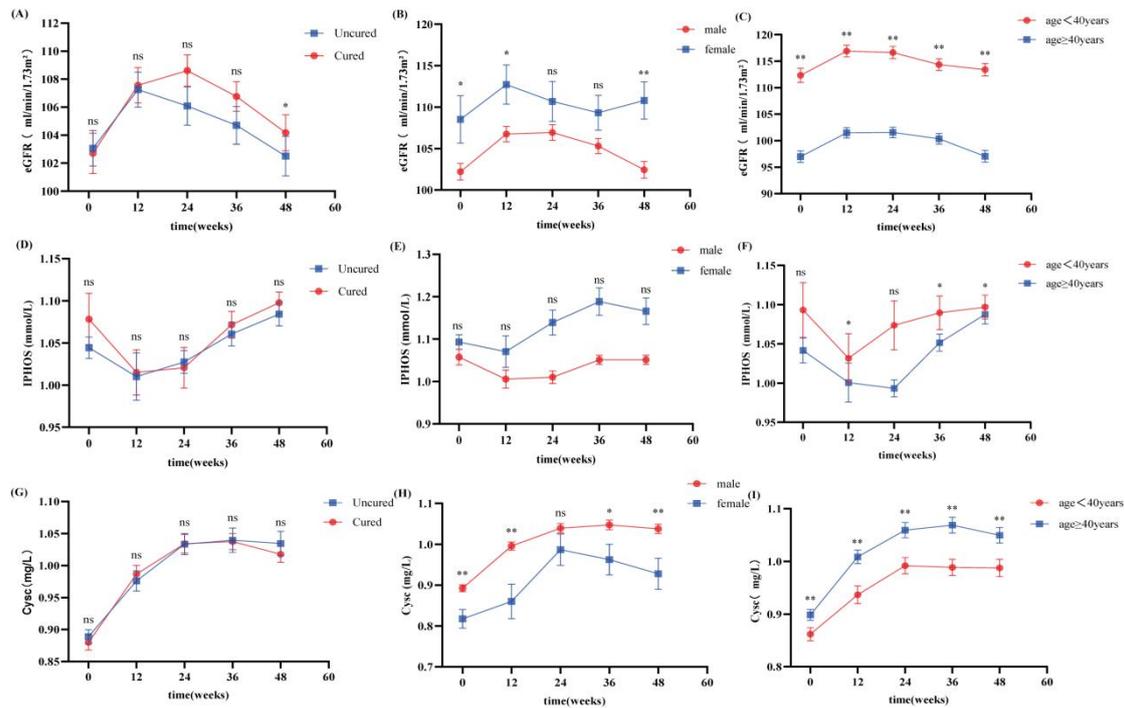

**FIGURE 2**

The renal function between different healing, gender and age states. (A) Changes in eGFR between different healing states. (B) Changes in eGFR between different genders. (C) Changes in eGFR at different ages (< 40 years, ≥ 40 years). (D) Changes in IPHOS between different healing states. (E) Changes in IPHOS between different genders. (F) Changes in IPHOS at different ages (< 40 years old, ≥ 40 years old). (G) Changes of CysC between different cure states. (H) Changes of CysC between different genders. (I) Changes of CysC at different ages (< 40 years old, ≥ 40 years old). $^{ns}p>0.05, ^*p<0.05, ^{**}p<0.001$.

3.3 The correlation between ALT levels and eGFR

The correlation between ALT levels and eGFR was positive in both the uncured and cured groups at 12 weeks and 24 weeks (P < 0.001, r=0.680, r=0.422, Figure 5A). After 12 weeks in the uncured group, there was a change of 36 weeks in estimated glomerular filtration rate (eGFR). A positive correlation was observed (P < 0.001, r=0.526, Figure 5b), with a poor correlation in the cured group and no statistically significant difference. Similarly, in the uncured group, after 12 weeks, there was a change of 48 weeks in eGFR. A positive correlation was found (P < 0.001, r=0.555, Figure 5C), with a poor correlation in the cured group and no statistical significance.

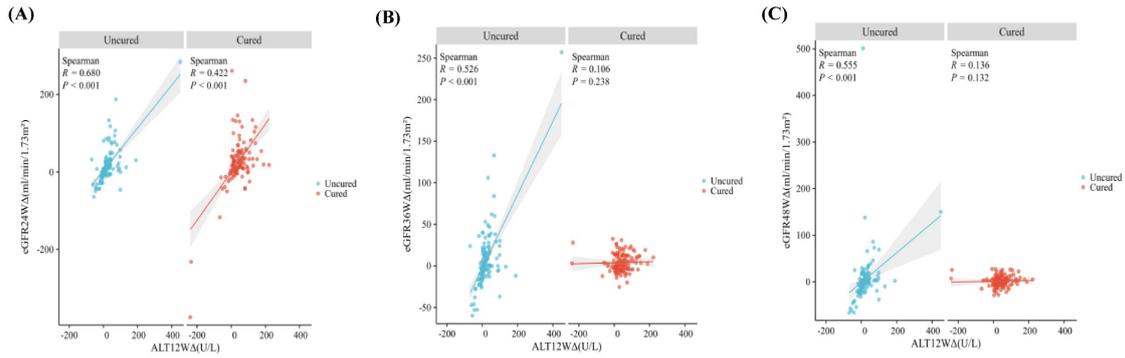

FIGURE 3

The correlation between ALT levels and eGFR.(A) ALT Δ at 12 weeks of treatment With eGFR Δ 24 weeks.(B) ALT Δ at 12 weeks of treatment With 36 weeks .(C) ALT Δ at 12 weeks of treatment With eGFR Δ 48 weeks Correlation analysis.ALT, alanine aminotransferase; Δ , Delta; eGFR, estimated glomerular filtration rate. Web page analysis tools from https://www.xiantao.love.

3.4 The renal function of various treatment regimens.

There were no statistically significant differences observed in the age and sex distributions among patients receiving pegylated interferon in combination with entecavir (ETV), pegylated interferon in combination with tenofovir disoproxil fumarate (TDF), and pegylated interferon monotherapy.The analysis of eGFR levels indicated that the three treatment regimens exhibited similar patterns of change over the 48-week treatment period, with no statistically significant differences observed (all $P > 0.05$). Similarly, the levels of IPHOS demonstrated consistent trends among the three treatment groups throughout the 48-week treatment duration, with no statistically significant variances at 0, 12, 24, and 36 weeks (all $P > 0.05$). However, a statistically significant difference was noted at the 48-week mark ($P = 0.013$).The levels of CysC indicated that the three treatment regimens exhibited a consistent pattern of change over a 48-week treatment period, with no statistically significant differences observed at 0, 12, 24, and 48 weeks (all $P > 0.05$). However, a significant difference was noted at 36 weeks ($P = 0.029$).

TABLE2. Participant' baseline characteristics after propensity-score matching.

| Baseline Characteristics | IFN-ETV (N=208) | IFN-TDF (N=208) | IFN alone (N=104) | P value |
|---|---|---|---|---|
| Age (year) | 36.1±9.1 | 35.3±9.2 | 36.6±8.8 | 0.462 |
| Sex | | | | 0.955 |
| male | 176 | 178 | 89 | |
| female | 32 | 30 | 15 | |
| Number of IFN | 47.31±18.79 | 47.90±16.98 | 50.89±21.31 | 0.261 |
| HBsAg (iu/ml) | | | | |
| 0 w | 418.33±366.09 | 444.98±477.61 | 344.93±295.55 | 0.117 |
| 12 w | 278.93±374.78 | 329.81±469.90 | 245.42±332.25 | 0.188 |
| 24 w | 170.05±300.37 | 207.46±360.38 | 192.31±345.52 | 0.520 |
| 36 w | 134.71±260.41 | 168.98±330.66 | 288.64±1515.15 | 0.205 |
| 48 w | 131.54±238.79 | 174.70±495.13 | 122.08±225.05 | 0.353 |
| Biochemistry data (renal function) | | | | |
| eGFR (ml/min/1.73m$^2$) | | | | |
| 0 w | 103.44±14.74 | 104.34±16.03 | 106.97±15.06 | 0.156 |
| 12 w | 107.54±13.57 | 109.59±14.01 | 111.29±14.52 | 0.067 |
| 24 w | 107.98±12.87 | 110.06±14.28 | 108.99±13.79 | 0.297 |
| 36 w | 106.17±13.20 | 108.96±12.84 | 107.16±15.51 | 0.107 |
| 48 w | 105.06±14.08 | 106.51±13.92 | 105.17±15.02 | 0.537 |
| IPHOS (mmol/L) | | | | |
| 0 w | 1.04±2.67 | 1.05±0.18 | 1.08±0.33 | 0.384 |
| 12 w | 1.00±0.26 | 1.05±0.39 | 1.00±0.23 | 0.258 |
| 24 w | 1.02±0.20 | 1.05±0.33 | 1.06±0.30 | 0.466 |
| 36 w | 1.07±0.20 | 1.08±0.16 | 1.09±0.35 | 0.641 |
| 48 w | 1.09±0.15 | 1.11±0.15 | 1.05±0.18 | 0.013 |
| Cysc (mg/L) | | | | |
| 0 w | 0.87±0.13 | 0.88±0.14 | 0.85±0.16 | 0.304 |
| 12 w | 0.97±0.16 | 0.95±0.16 | 0.94±0.16 | 0.191 |
| 24 w | 1.01±0.16 | 1.01±0.16 | 1.00±0.19 | 0.875 |
| 36 w | 1.03±0.17 | 1.00±0.15 | 0.98±0.20 | 0.029 |
| 48 w | 1.01±0.78 | 0.99±0.16 | 1.00±0.17 | 0.373 |

**Abbreviation**: IFN-ETV, pegylated interferon entercavri, pegylated interferon combined with entecavir; IFN-TDF, pegylated interferon tenofovir disoproxil, pegylated interferon combined with tenofovir fumarate; IFN, pegylated interferon, pegylated interferon

3.5 Baseline characteristics of various stages of renal function.

Based on the eGFR level at 48 weeks of treatment, patients were categorized into stage 1group and stage 2 group.
The study found significant age differences between the two groups, with the stage 1 group being significantly older than the stage 2 group (P < 0.001). There was no significant difference in the use of pegylated interferon (P = 0.452).
Baseline levels of eGFR, IPHOS, and CysC showed significant differences (all P < 0.05).eGFR and CysC levels showed significant differences over 48 weeks of treatment (P < 0.001), while IPHOS levels did not differ significantly between the two groups at 12, 24, 36, and 48 weeks (all P > 0.05).

TABLE3.Characteristics of participant with different renal function stage.

| Baseline Characteristics | Renal function (stage1) (N=73) | Renal function (stage2) (N=447) | P-value |
|---|---|---|---|
| Age | 41.60±8.80 | 34.90±8.80 | <0.001 |
| Number of IFN | 48.60±16.72 | 48.21±18.96 | 0.452 |
| HBsAg (iu/ml) | | | |
| 0W | 362.63±365.66 | 422.75±409.54 | 0.234 |
| 12W | 276.23±371.15 | 295.25±414.89 | 0.842 |
| 24W | 213.59±366.87 | 185.53±328.96 | 0.236 |
| 36W | 176.57±341.80 | 179.63±341.80 | 0.257 |
| 48W | 159.59±316.37 | 144.84±369.29 | 0.477 |
| Biochemistry data | | | |
| AST (U/L) | | | |
| 0W | 27.55±9.12 | 28.08±14.64 | 0.185 |
| 12W | 55.68±41.35 | 58.46±45.35 | 0.444 |
| 24W | 52.48±48.36 | 52.80±33.64 | 0.327 |
| 36W | 41.24±19.85 | 43.61±26.16 | 0.844 |
| 48W | 34.72±17.16 | 41.84±80.71 | 0.913 |
| ALT (U/L) | | | |
| 0W | 30.29±14.09 | 31.62±20.42 | 0.471 |
| 12W | 63.62±62.01 | 68.44±45.81 | 0.084 |
| 24W | 54.25±65.32 | 60.93±58.19 | 0.016 |
| 36W | 42.73±19.47 | 47.34±38.32 | 0.558 |
| 48W | 45.33±65.76 | 44.68±60.80 | 0.560 |
| eGFR (ml/min/1.73m²) | | | |
| 0W | 87.48±12.92 | 107.29±13.87 | <0.001 |
| 12W | 92.06±11.52 | 111.90±12.28 | <0.001 |
| 24W | 92.10±11.49 | 111.78±11.86 | <0.001 |
| 36W | 89.57±12.08 | 110.41±11.41 | <0.001 |
| 48W | 81.01±7.67 | 109.68±10.46 | <0.001 |
| IPHOS (mmol/L) | | | |
| 0W | 1.03±0.25 | 1.06±0.25 | 0.009 |
| 12W | 1.01±0.20 | 1.02±0.25 | 0.104 |
| 24W | 1.00±0.20 | 1.04±0.29 | 0.103 |
| 36W | 1.09±0.29 | 1.08±0.21 | 0.256 |
| 48W | 1.09±0.15 | 1.09±0.16 | 0.947 |
| Cysc (mg/L) | | | |
| 0W | 0.94±0.14 | 0.86±0.14 | <0.001 |
| 12W | 1.08±0.18 | 0.94±0.15 | <0.001 |
| 24W | 1.11±0.19 | 0.99±0.16 | <0.001 |
| 36W | 1.13±0.19 | 0.99±0.16 | <0.001 |
| 48W | 1.12±0.17 | 0.98±0.16 | <0.001 |

3.6 Variation tendency of various stages of renal function.

Participant were divided into stage 1 and stage 2 groups based on their baseline renal function. Renal function in both groups increased and then decreased at 48 weeks, with higher eGFR levels in the IFN-TDF group compared to the other two groups. There was no significant difference in renal function between stage 1 and stage 2 groups, but there was a significant difference in eGFR at 12, 24, and 36 weeks in the stage 2 group ($P < 0.05$).

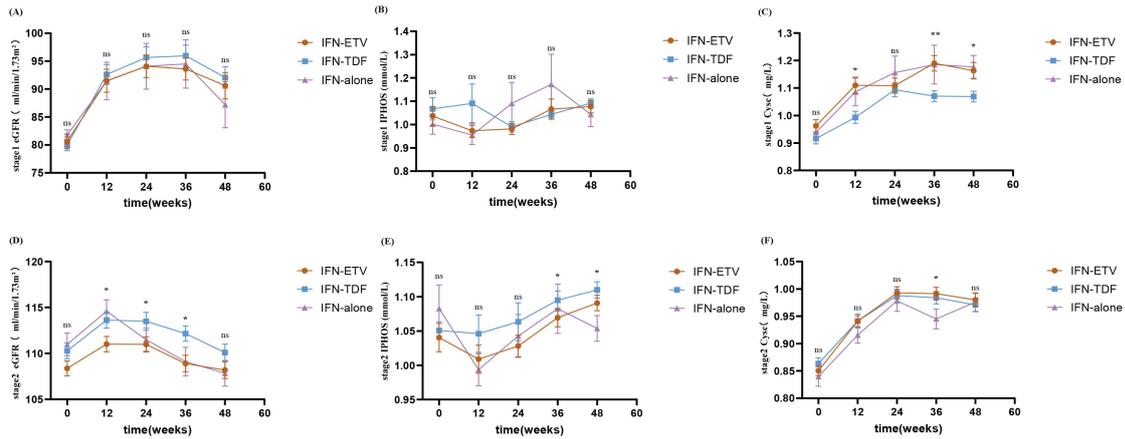

**FIGURE 4**
Variation tendency of various stages of renal function.
(A).Variation tendency of stage 1 of eGFR.(B).Variation tendency of stage 1 of IPHOS.(A).Variation tendency of stage 1 of Cysc.(A).Variation tendency of stage 2 of eGFR.(A).Variation tendency of stage 2 of IPHOS.(A).Variation tendency of stage 2 of Cysc.

## 3.7 Variation of various stages of renal function.

Based on baseline renal function, patients were split into stage 1 and stage 2 groups. Renal function changes were monitored at 12, 24, 36, and 48 weeks. In the stage 1 group, only CysC at 36 weeks showed statistically significant changes with one treatment regimen ($P < 0.05$). Other changes were not significant ($P > 0.05$). In stage 2 group, only eGFR at 36 and 48 weeks, and IPHOS at 48 weeks showed statistically significant changes in renal function among the three treatment regimens ($P < 0.05$). No other differences were significant ($P > 0.05$).

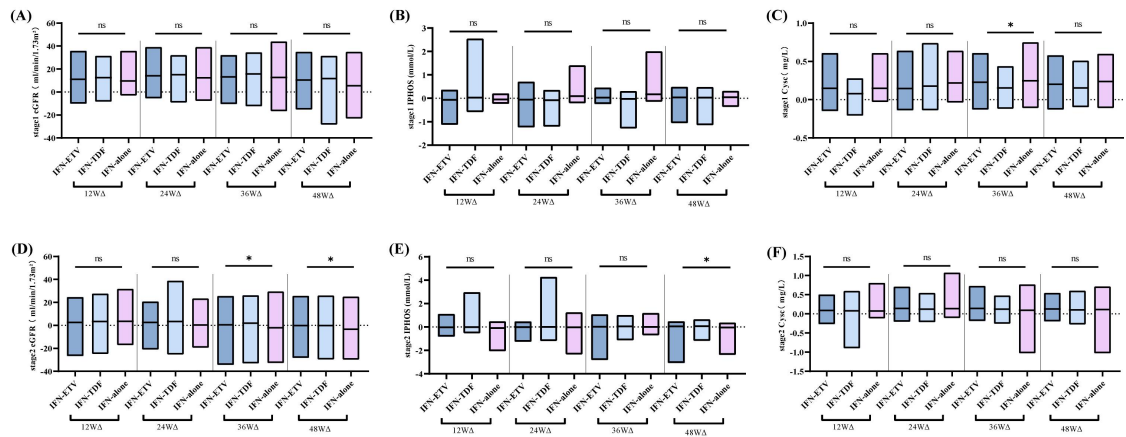

**FIGURE 5**
Variation of various stages of renal function.
(A).Variation of stage 1 of eGFR.(B).Variation of stage 1 of IPHOS.(A).Variation of stage 1 of Cysc.(A).Variation of stage 2 of eGFR.(A).Variation of stage 2 of IPHOS.(A).Variation of stage 2 of Cysc.

## 3.8 Analysis of factors predicting renal outcomes at 48 weeks

Lasso regression found that 19 variables were initially included in the equation, but only 8 variables were ultimately effective based on cross validation results.

Multivariate analysis in the matched cohort of the three treatment regimens revealed that age, the number of interferons used, baseline HBsAg level, baseline AST level, baseline ALT level, baseline IPHOS level, and baseline CysC level did not exhibit statistical significance in predicting the 48-week renal outcome (all $P > 0.05$). Conversely, baseline eGFR level emerged as a significant predictor of renal outcome at 48 weeks ($P = 0.017$).

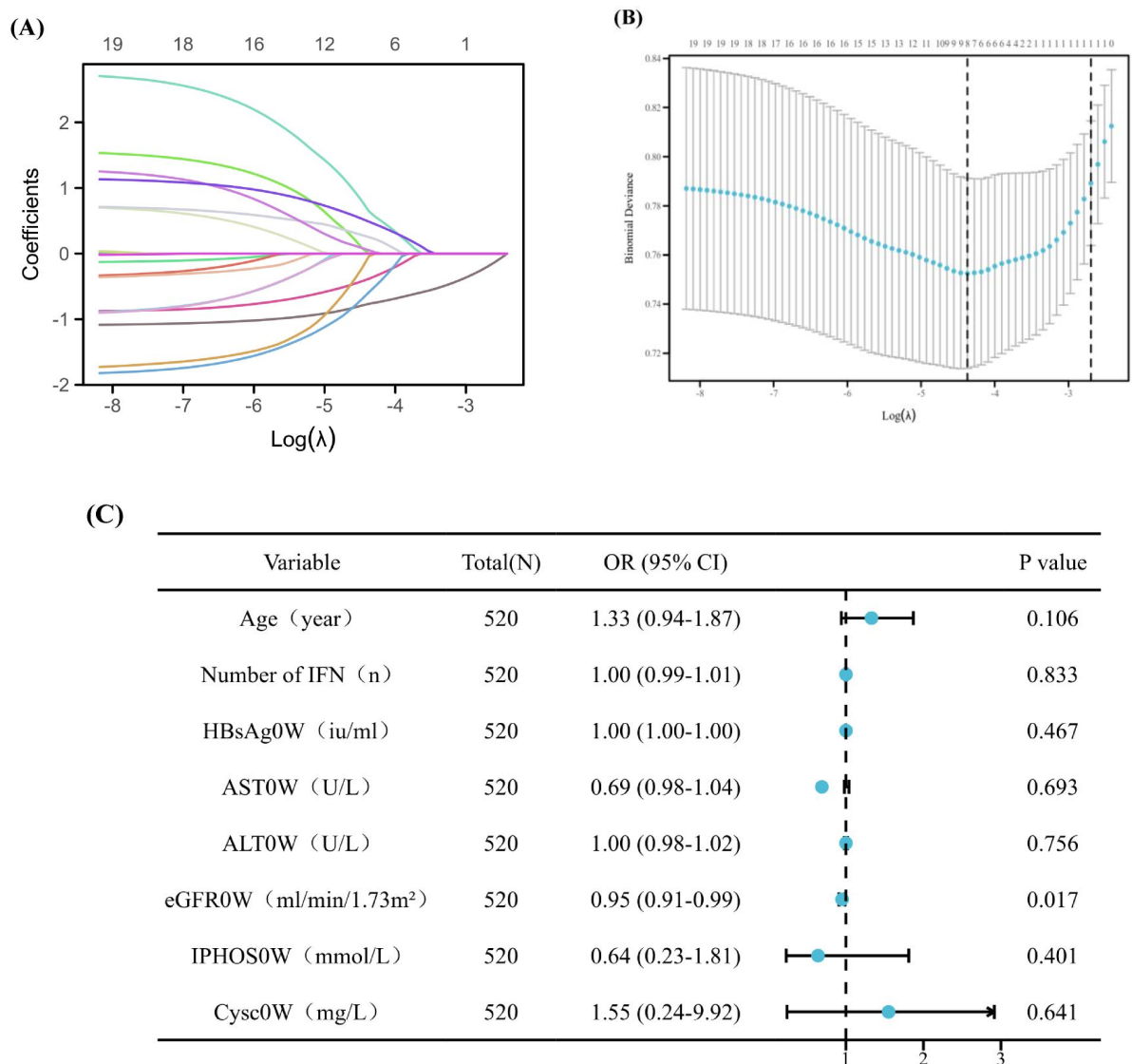

**FIGURE 6**

Analysis of factors predicting renal outcomes at 48 weeks.
(A). Lasso coefficient path diagram .(B). Lasso regression analysis cross validation song.(C). multivariate analysis forest plot of 48 weeks renal outcomes in participant.

## 3.9 Efficacy test of predictors for renal outcome.

Four factors were chosen for predictive testing based on lasso regression and multivariate analysis. The first two factors, baseline eGFR and age, were selected using the AUC value. Logistic regression was used to predict eGFR outcomes at 48 weeks in patients. The area under the receiver operating characteristic curve (AUC) of the prediction model is 0.851 (95% CI: 0.807, 0.895). At a prediction probability threshold of 0.853, the Youden index is maximized for patients with normal renal function at 48 weeks of treatment, yielding a sensitivity and specificity of 0.781 each. We created nomograms and analyzed a calibration curve to evaluate the impact of gender, age, cure status, and eGFR level on renal function after 48 weeks of treatment. The calibration curve had a C index of 0.729, showing moderate accuracy.

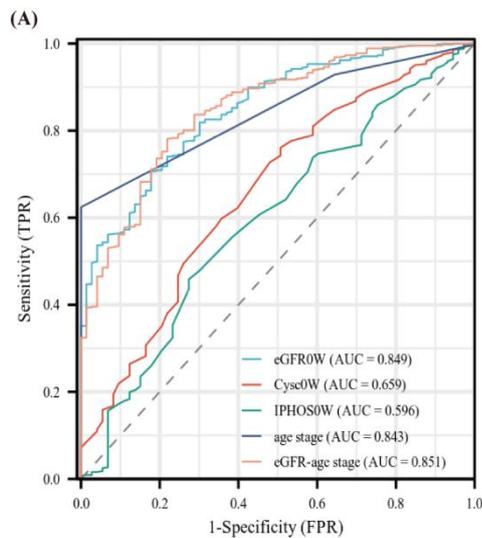

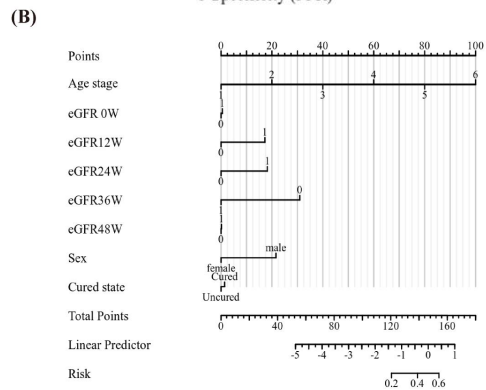

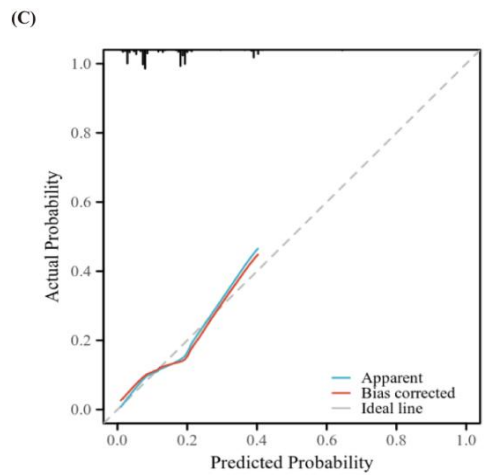

**FIGURE 7**

Efficacy test of predictors for renal outcome.

(A).ROC curve for predicting renal outcome at 48 weeks of treatment.(B). Nomogram for predicting renal outcome at 48 weeks of treatment.(C). Calibration curve for predicting renal outcome at 48 weeks of treatment (500 replicates).

## 4 DISCUSSION

The objective of this empirical investigation is to enhance the likelihood of clinical remission and prognosis for a greater number of individuals affected by chronic hepatitis B. By augmenting the remission rates of chronic hepatitis B and diminishing the population of hepatitis B carriers, it underscores the imperative for the nation to mitigate "morbidity and mortality." The establishment of a substantial cohort receiving interferon treatment for chronic hepatitis B aims to identify optimal patient profiles and treatment protocols conducive to clinical remission through the analysis of extensive datasets, thereby facilitating their widespread implementation.Enhance doctors' ability to treat chronic hepatitis B through specialized training.It offers long-term renal function safety data for patients receiving pegylated interferon or oral antiviral drugs, specifically looking at renal function injury. This data is crucial for understanding the drugs' impact on renal function in clinical trials and cohort studies. Patients who were clinically cured at 48 weeks had better renal function compared to those who were not cured, indicating a better response to pegylated interferon and potential benefits in renal function. This aligns with previous studies suggesting that pegylated interferon therapy may improve eGFR in patients[13].There was no notable difference in renal function outcomes after 48 weeks between patients treated with pegylated interferon combined with ETV and TDF, suggesting oral antiviral drugs may be a suitable choice for CHB patients undergoing interferon combined therapy.First-line nucleoside drugs are primarily excreted in urine, with TDF showing dose-dependent nephrotoxicity in both animal and human studies. TDF is a water-soluble diester prodrug of tenofovir with a diphenylfuroate group, but its poor stability and low utilization are due to its short half-life and inability to selectively decompose and release in the liver[15].As TDF is taken for longer periods, the number of reported kidney side effects is rising[16].Koklu et al discovered that TDF is more effective at reducing eGFR levels compared to other nucleoside drugs[17]. Observational studies suggest that TDF may increase the risk of renal function injury, especially in older or menopausal patients. There is still debate about treating renal function with interferon alone or in combination with oral antiviral drugs.

Numerous studies have demonstrated that interferon can enhance kidney function in patients with renal insufficiency and transplants[19,20,21,22].The exact way interferon protects the kidneys is still unknown. This study suggests that pegylated interferon's high content and shape may slow down drug excretion in the kidneys, potentially protecting them.Studies have found that lower levels of peripheral $CD^4CXCR^5$ follicular helper T cells are linked to higher eGFR values, indicating a potential role in improving eGFR through IFN[23].There are limited studies on how interferon or combined oral antiviral drugs affect patients' kidney function.Prior research found[24] no significant difference in average eGFR between TDF and ETV in patients with normal renal function. However, in patients with moderate renal impairment, the average eGFR was lower in the TDF group and decreased during treatment. The study also noted that viral inhibition could improve renal function to some extent, potentially explaining why combined treatments showed no significant difference.

Our study offers clinicians a more precise selection of antiviral drug regimens by using propensity score matching to balance patient characteristics.The findings from the observation of renal function in two groups, one uncured and the other cured, undergoing a 48-week treatment with interferon combined with oral antiviral drugs, indicated that the renal function of the cured group was superior to that of the uncured group. Furthermore, the similar change trend observed in both groups indirectly suggests a potential association between the response to pegylated interferon and renal function outcomes in patients.

Subsequently, propensity score matching was employed to mitigate any imbalances in clinical characteristics among the three cohorts receiving pegylated interferon alone, pegylated interferon in conjunction with ETV, and pegylated interferon in conjunction with TDF. The study found no discernible variance in the impact of the three treatment protocols on renal function outcomes after a 48-week period.Furthermore, the findings suggest that TDF offers a wider range of treatment options in combination therapy compared to ETV or oral antiviral drugs alone. Additionally, the high quality of the study cohort allowed for the accurate prediction of renal function outcomes at 48 weeks. Age and baseline eGFR levels were identified as significant predictors in assessing renal outcomes, underscoring the importance for clinicians to promptly monitor renal function in older patients or those with poor baseline renal function.

This study is subject to certain limitations. The disease type prevalent in the study cohort is predominantly singular, thus limiting its ability to accurately represent the complexities present in patients with renal complications. Additionally, the duration of observation in this study is relatively short, suggesting that further research with an extended observation period may be necessary for comprehensive validation.

**AUYHOR CONTRIBUTIONS**

Jinhua Zhao and Lili Wu had full access to all the data in the study and take responsibility for the integrity of the data and the accuracy of the data analysis.Study concept and design: Hong Deng and Zhiliang Gao.Acquisition,analysis or interpretation of data:All authors. Drafting of the manuscript:Jinhua Zhao and Lili Wu. Critical revision of the manuscript for important intellectual content: All authors. Statistical analysis:Jinhua Zhao、Lili Wu、Xiaoan Yang. Obtained funding:Hong Deng and Zhiliang Gao. Administrative, technical or material support: Hong Deng and Zhiliang Gao. Study supervision: Hong Deng.        .

**ACKNOWLEDGEMENTS**
We would like to thank the participants and staff of the Third Affiliated Hospital of Sun Yat-sen University for their valuable contributions.

## FUNDING INFORMATION

This work was supported by the National Natural Science Foundation of China ,Grant Number:81870597,82170612; the Guangzhou Science and Technology Program Key Projects,Grant Number:2023B01J1007. The funders had no role in design and conduct of the study; collection, management, analysis and interpretation of the data; and preparation, review or approval of the manuscript.

## CONFLICT OF INTEREST STSTEMENT

The authors declare that they have no conflicts of interest.

## ETHICAL STATEMENT

The study protocol was consistent with the International Conference on Harmonization Guidelines, applicable regulations, and the ethical guidelines of the Declaration of Helsinki. The protocol and consent forms were approved by the Research Ethics Committee of the Third Affiliated Hospital of Sun Yat-sen University, China ([2018]02-218-02). Patients with antiviral therapy were obtained from a real-world study registered at clinicaltrails.gov (chictr180020369).

## DATA SHARING STATEMENT

The datasets used in the study are available from the corresponding author on reasonable request.

## ORCID

JInhua Zhao https://orcid.org/0009-0008-1757-8841
Lili Wu https://orcid.org/0000-0002-1319-7274